\documentstyle[12pt]{article}
\advance\hoffset by -7mm  
\makeatletter
\@addtoreset{equation}{section}
\makeatother

\renewcommand{\title}[1]{\null\vspace{-10mm}

\noindent{\begin{center}\normalsize{\bf #1}\end{center}}\vspace{5mm}

\noindent {By }}
\newcommand{\authors}[1]{\noindent{#1}\vspace{0mm}

}
\renewcommand{\abstract}[1]{\vspace{5mm}

\noindent{\small{\em Abstract.} #1}\vspace{0mm}

}

\begin{document}
 
\newcommand{\andy}[1]{ }
\renewcommand{\thesection}{{\normalsize \arabic{section}}} 

\title{NON-INVARIANT VELOCITY OF LIGHT AND CLOCK SYNCHRONISATION IN 
ACCELERATED SYSTEMS}
\authors{F. Goy\footnote{Financial support of the Swiss National Science 
Foundation and the Swiss Academy of Engineering Sciences.} - Dipartimento di 
Fisica - Universit\^^ {a} di Bari - Via G. Amendola, 173 - 70126  Bari - Italy  
 - E-mail:  goy@axpba1.ba.infn.it}
 
\abstract{
Clock synchronisation is conventional when inertial systems are involved. This 
statement is no longer true in accelerated systems. A demonstration is given in
the case of a rotating platform. We conclude that theories based on the 
Einstein's clock synchronisation procedure are unable to explain, for example, 
the Sagnac effect on the platform. Implications on very precise clock 
synchronisation on earth are discussed.}
 
\section{{\small{\sc Conventionality of clock synchronisation
 in inertial frames}}}
\noindent Let us begin with an operationalistic definition of physical time:
``time is what is measured by clocks''. If we want to apply this definition for 
the setting of clocks not only 
in one point, but everywhere in space, we see that two
different and independent notions are implied:
\begin{itemize}
\item The rate at which time flows in each point.
\item The simultaneity of events in different points of space.
\end{itemize}
Following Poincar\'{e} \cite{poin:04a} and Einstein \cite{eins:05a}, we can use 
light signals for the setting of two clocks  of identical fabrication, which 
are at two points $A$ and $B$ of an inertial frame. We send a light signal from 
$A$ at time $t_{1}$, which arrives at $B$ at time $t_{2}$ and comes back at $A$ 
at time $t_3$. The time $t_{2}$ of $B$ is {\em defined} to be synchronous 
with the midtime of departure and arrival in $A$. This definition is called the 
Einstein's synchronisation. Mathematically:
\andy{biere}
\begin{equation}
t_{2} = t_{1} + \frac{1}{2}(t_{3} - t_{1})
\label{eq:biere}
\end{equation}
Reichenbach commented \cite{reic:58a}:{\em "This definition is essential for 
the special theory of relativity, but not epistemologically necessary. If we 
were to follow an arbitrary rule restricted only to the form}
\andy{bibine}
\begin{equation}
t_{2} = t_{1} + \varepsilon(t_{3} - t_{1}) \;\;\;\; 0 < \varepsilon < 1
\label{eq:bibine}
\end{equation}
{\em it would likewise be adequate and could not called be false. If the 
special theory of relativity prefers the first definition, i.e., sets 
$\varepsilon$ equal to $1/2$, it does so on the ground that this definition 
leads 
to simpler relations.}" On the possibility to choose freely $\varepsilon$ 
according 
to (\ref{eq:bibine}) agreed, among others, Winnie \cite{winn:70a}, Gr\"{u}nbaum 
\cite{grun:73a}, Jammer 
\cite{jamm:79a}, Mansouri and Sexl \cite{mase:77a}, Sj\"{o}din \cite{sjod:79a},
Cavalleri and Bernasconi \cite{cava:89a}, Ungar \cite{unga:91a}, 
Vetharaniam and Stedman 
\cite{vest:91a}, Anderson and Stedman \cite{anst:92a}, \cite{anst:94a}. 
Clearly, 
different values of $\varepsilon$ correspond to different 
values of the one way-speed of light. 

\noindent A slightly different position was developed in the parametric test 
theory of special relativity of Mansouri and Sexl. Following these authors, 
we assume  that there is {\em at least one} inertial frame in which light 
behaves isotropically. We call it the priviledged frame $\Sigma$ and denote 
space and time coordinates in this frame by the letters: $(x_{0},y_{0},z_{0},
t_{0})$. 
In $\Sigma$, clocks are synchronised with Einstein's 
procedure.
We consider also an other system $S$ moving with uniform 
velocity $v<c$ along 
the $x_{0}$-axis in the positive direction. In $S$, the coordinates are written 
with 
lower case letters $(x,y,z,t)$.
Under rather general assumptions, symmetry conditions on the two systems, 
the assumption that the two-way velocity of light is $c$ and furthermore
that the time 
dilation factor has its relativistic value, one can derive
the following transformation:
\andy{stp}
\begin{eqnarray}
x & = & \frac{1}{\sqrt{1-\beta^{2}}}\left(x_{0}-vt_{0}\right) \nonumber \\
y & = & y_{0} \nonumber \\
z & = & z_{0} \label{eq:stp} \\
t & = & s\left(x_{0}-vt_{0}\right) + \sqrt{1-\beta^{2}}\;\; t_{0}\;\;, \nonumber
\end{eqnarray}
where $\beta=v/c$. The parameter $s$, which determines the synchronisation 
in the $S$ frame remains unknown. Einstein's synchronisation 
in $S$ involves: 
\mbox{ $s=-v/(c^{2}\sqrt{1-\beta^{2}})$} and (\ref{eq:stp}) becomes a Lorentz 
boost. For a general $s$, the inverse one-way velocity of light is given
by \cite{sell:94a}:
\andy{lexo}
\begin{equation}
\frac{1}{c_{\rightarrow}(\Theta)} = \frac{1}{c} + \left(\frac{\beta}{c} +
s\sqrt{1-\beta^{2}}\right)\cos\Theta\;\;,
\label{eq:lexo}
\end{equation}
where $\Theta$ is the angle between the $x$-axis and the light ray in $S$.
$c_{\rightarrow}(\Theta)$ is in general dependent on the direction. 
A simple case is $s=0$. 
This means from (\ref{eq:stp}), that at $T=0$ of $\Sigma$ we set all clocks 
of $S$ 
at $t=0$ (external synchronisation), or that we synchronise the clocks by means 
of light rays with velocity $c_{\rightarrow}(\Theta)=c/(1+\beta\cos\Theta)$
(internal synchronisation).
It should be stressed that, unlike to the parameters of length contraction 
and time dilation, {\em the parameter s cannot be tested}, but its 
value must be 
assigned in accordance with the synchronisation choosen in the experimental 
setup. It means, as regards experimental results, that theories using 
different s are equivalent. Of course, they may predict different values of
physical quantities (for example the one-way speed of light). This 
difference resides not in nature itself but in the convention used for the 
synchronisation of clocks.
For a recent and comprehensive discussion of this subject, see 
\cite{vest:93a}. A striking consequence of (\ref{eq:lexo}) is that the negative 
result 
of the Michelson-Morley experiment does not rule out an ether. Only an ether 
with galilean transformations is excluded, because the galilean transformations 
do not lead to an invariant two-way velocity of light in a moving system.

\noindent Strictly speaking, the conventionality of clock synchronisation was 
only shown to hold in inertial frames. The derivation of equation 
(\ref{eq:stp}) is done in inertial frames and is based on the assumption that 
the two-way velocity of light is constant in all directions. This last 
assumption is no longer true in accelerated systems. But special relativity is 
not only used in inertial frames. A lot of textbooks bring examples 
of calculations done in accelerated systems, using infinitesimal  Lorentz 
transformations. Such calculations use an additional assumption: the so-called 
{\em Clock Hypothesis}, which states that the rate of an accelerated ideal 
clock is identical to that of the instantaneously comoving inertial frame.  
This hypothesis first used implicitely by Einstein in his article of 1905 was 
superbly confirmed in the famous timedecay experiment of muons in the CERN, 
where the muons had an acceleration of $10^{18}g$, but where their timedecay 
was only due to their velocity \cite{bail:77a}. We stress here 
the logical independence of this assumption from the structure of special 
relativity as well as from the assumptions necessary to derive (\ref{eq:stp}). 
The opinion of the author is that the {\em Clock Hypothesis}, added to special 
relativity 
in order to extend it to accelerated systems leads to logical contradictions 
when the question of synchronisation is brought up. This idea was also 
expressed by Selleri \cite{sell:96a}.
The following example (see \cite{mast:93a}) shows it: imagine that 
two distant clocks are screwed on an inertial frame (say a train) and 
synchronised with an 
Einstein's synchronisation. The train accelerates 
during a certain time. After that, the acceleration stops and the train has 
again an inertial mouvement. It is easy to 
show that the clocks are no more Einstein's synchronous. {\em So the 
Clock 
Hypothesis is inconsistent with the clock setting of relativity}. On the
other hand, the {\em Clock Hypothesis} is tested with a high degree of 
accuracy \cite{eisl:87a} 
and cannot be rejected, so one has to reject the clock setting of special 
relativity. The only 
theory which is consistent with the {\em Clock Hypothesis} is based on 
transformations (\ref{eq:stp}) with $s=0$. This is an ether theory. 
The fact that only an ether 
theory is consistent with accelerated motion give strong evidences that an 
ether exist, but does not involve inevitably 
that our velocity relative to the ether is measurable. It remains an important 
open 
question which is beyond the scope of this paper. The opinion of the author is 
that it cannot be measured in the above example, in spite of the logical 
difficulties of special relativity. There are strong evidences that when a 
temporary
acceleration is used to go from one inertial frame to an other, an ``absolute'' 
motion cannot be measured, as it has been shown by the author in the case of 
stellar aberration \cite{fgoy:96a}. The situation could be different when 
a permanent acceleration acts on a system. 
\noindent The logical difficulties of special relativity indicates that the 
relativity principle (The law of physics are the same in all inertial 
frames)
does not only express an objective property of nature, but also a human choice 
about the setting of clocks. The objective part of the relativity principle is 
expressed in the statement: an uniform motion through the ether is not 
detectable. Equations (\ref{eq:stp}) obeys to this second statement.  

\section{{\small{\sc The Sagnac effect}}}
\noindent An uniform motion through the ether was never detected, what is 
expressed in the negative result of the Michelson-Morley experiment. In 
contrary, rotational motion can be detected by mechanical methods like the 
Foucault 
pendulum as well as by luminuferous methods  like in the Sagnac effect. Eight 
years after 1905, Sagnac wrote an article whose title shows that he thought to 
have proved the reality of ether \cite{sagn:13a}. The Sagnac effect 
is essentially the 
observation of the phase shift between two coherent beams travelling on 
opposite paths in an interferometer placed on a rotating platform.
In 1925, Michelson and Gale \cite{miga:25a}, detected the earth rotation rate
with a giant interferometer constructed with this aim. Michelson explained the 
measured effect with a velocity of light which is not constant. 
Nowdays the Sagnac effect is observed with light (in ring 
lasers and fiber optics interferometers \cite{anbi:94a}) and in interferometers 
built for electrons \cite{hani:93a}, neutrons \cite{coov:75a}, atoms 
\cite{stco:94a} and superconducting Cooper pairs \cite{fach:96a}. The phase 
shift in interferometers is a consequence of the time delay between the arrival 
of the two beams, so a Sagnac effect is also measured directly with atomic 
clocks timing light beams sent around the earth via satellites \cite{alal:85a}.
 
\noindent In the typical experiment for the  study of the effect, a 
monochromatic light 
source placed on a disk emits two coherent light beams in opposite directions 
along the circumference until they reunite after a $2\pi$ propagation. The 
positionning of the interference figure depends on the disk rotational 
velocity. Textbooks deduce the Sagnac formula 
in the laboratory, but say nothing about 
the description of the phenomenon on the rotating platform. Exception to this 
trend are Michelson \cite{miga:25a},
 Langevin \cite{lang:21a}, Anandan \cite{anan:81a}, Dieks and Nienhuis 
\cite{dini:90a} and Post \cite{post:67a}, but dissatisfaction remains 
widespread, because none of these treatments is free of ambiguities. Michelson
uses a noninvariant velocity of light, but with a galilean addition of 
velocities, which is in contradiction with his experiment of 1887. 
Langevin, in his 1937, paper recognized the possibility of a nonstandard 
velocity of light on the rotating platform, but gives a formula only valid to 
the first order. Post's relativistic formula is not generally valid, but 
limited to the case where the origin of the tangential inertial frame coincides 
with the center of the rotating disk.

\noindent So let us deduce here the Sagnac formula 
for a circular Sagnac device rotating
in the anticlockwise sense in the priviledged frame $\Sigma$, for simplicity. 
We first
calculate the time difference between a clockwise 
and an anticlockwise beam of light constrained to follow a circular path of 
radius R as seen by an observer in the laboratory and secondly 
we will make the calculation on the rim of the disk. The two 
beams leave the beamsplitter at time $t_{0}=0$  when it is in position C
(see figure 1). 
The clockwise
circulation is opposite to the direction of rotation and meets the 
beamsplitter when it is in position C' shifted by $\Delta s'$ with respect 
to C, 
at time $t_{0}'$. 
The anticlockwise beam, travelling in the same direction as the direction of 
rotation meets the beamsplitter in the later position C'', shifted by $\Delta 
s''$, with respect to C at time $t_{0}''$. The geometry is given in the 
following figure. $\omega$ is the rate of rotation of the interferometer.
\vspace{-20pt}
        \begin{figure}[ht]
        \let\picnaturalsize=N
        \def\picsize{6.0cm}
        \def\picfilename{lfig1.eps}
        \ifx\nopictures Y\else{\ifx\epsfloaded Y\else\input epsf \fi
        \let\epsfloaded=Y
        \centerline{\ifx\picnaturalsize N\epsfxsize \picsize\fi
        \epsfbox{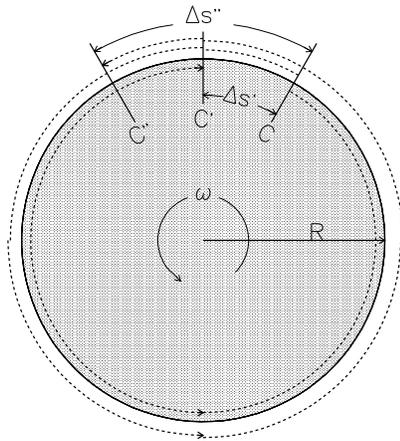}}}\fi
\caption{Simplified Sagnac configuration: the light beams are drawn with dashed 
lines}
\end{figure}

\noindent We have:
\andy{bleue}
\begin{eqnarray}
\Delta s' = \omega R t_{0}'\;\;;\;\;\;& & L_{0}-\Delta s'= c t_{0}'
\label{eq:bleue}
\end{eqnarray}
and
\andy{absinthe}
\begin{eqnarray}
\Delta s'' = \omega R t_{0}''\;\;;\;\;\;& & L_{0}+\Delta s''= c t_{0}''\;\;,
\label{eq:absinthe}
\end{eqnarray}
where $L_{0}$ is the circumference of the rotating disk measured in the 
laboratory.
The length of this circumference is reduced relative to the length 
$L=2\pi R$ measured directly on the disk (see section 3 for a discussion of the 
geometry of the rotating disk). We have: 
$L_{0}=L\sqrt{1-\omega^{2}R^{2}/c^{2}}$. 
Eliminating $\Delta s'$ and $\Delta s''$ from equations (\ref{eq:bleue}) and 
(\ref{eq:absinthe}), we obtain
the time difference $\Delta t_{0}=t_{0}''-t_{0}'$ in the laboratory 
\andy{gueuse}
\begin{equation}
\Delta t_{0}=\frac{4\omega A}{c^{2} \sqrt{1-\omega^{2}R^{2}/c^{2}}}\;\;,
\label{eq:gueuse}
\end{equation}
where $A$ stands for $\pi R^2$.
Because of the time dilation, an observer on the platform must find a 
time difference 
$\Delta t$, so that:
\andy{cardinal}
\begin{equation}
\Delta t_{0}=\frac{\Delta t}
                          {\sqrt{1-\omega^{2}R^{2}/c^{2}}}\;\;,
\label{eq:cardinal}
\end{equation}
which gives using (\ref{eq:gueuse}):
\andy{heineken}
\begin{equation}
\Delta t=\frac{4\omega A}{c^2}
\label{eq:heineken}
\end{equation}

\noindent Let us now directly calculate the result on the platform, as done by
Selleri \cite{sell:96a}. Every infinitesimally small portion of the rim of the 
disk can be considered to be at rest in a comoving inertial frame 
tangential to the 
disk. The {\em Clock Hypothesis} of special relativity tells us that the rate 
of ideal clocks on the rim is the same as in the comoving inertial frame.
Similary ideal unit measuring rods behaves in the same way on the rim and in 
the comoving inertial frame \cite[p. 254]{moll:72a}. For reasons of 
continuity, we must choose the same synchronisation on a small portion of the 
rim as in the tangeantial inertial frame. The one-way velocity of light is 
given by (\ref{eq:lexo}). The time difference between the arrival of the two 
light beams as calculated on the disk is given by:
\andy{perroni}
\begin{eqnarray}
\Delta t = t''- t' = 
L\left(\frac{1}{c_{\rightarrow}(0)}-\frac{1}{c_{\rightarrow}(\pi)}\right)
=\frac{4\omega A}{c^2}\left(1+\frac{s c^{2} \sqrt{1-\omega^{2}R^{2}/c^{2}}}
  {\omega R}\right)
\label{eq:perroni}
\end{eqnarray}
Comparing this last formula to (\ref{eq:heineken}), we see that the only value 
giving a correct prediction of the Sagnac effect on the platform is $s=0$. In 
all other cases, the formula (\ref{eq:heineken}) is not recovered as it should 
be, because (\ref{eq:gueuse}) is predicted by all theories having the general 
transformation ({\ref{eq:stp}) since they all use an Einstein's 
synchronisation in the laboratory. In particular, the theory of special 
relativity gives $\Delta t = 0$, when the value of $s$ given after 
(\ref{eq:stp}) is substitued in (\ref{eq:perroni}).

Independently of the above considerations, we would draw the attention of the 
reader 
to our formula (\ref{eq:gueuse}), which differs in the second order in $\omega 
R/c$, from the formula given by Post \cite{post:67a} and followers. In the 
same physical situation, he gives in the laboratory: 
\andy{denner}
\begin{equation}
\Delta t_{0}^{Post}=\frac{4\omega A}
                         {c^{2} \left(1-\omega^{2}R^{2}/c^{2}\right)}\;\;,
\label{eq:denner}
\end{equation}
The discrepancy comes from the fact that Post does not take account of the 
Lorentz contraction of the circumference of the disk. From our point of view,
this is erroneous. Unfortunately, the 
precision of measures of the Sagnac effect does not enable to decide between 
the two formulas on an experimental level, so that we have no empirical 
information on the physics of a relativistic rotating disk. 

\section{{\small{\sc Clock synchronisation in the formalism of general 
relativity}}}

\noindent The same disk as in section 2 is considered. We can obtain the 
metric on the disk as follow: in the 
inertial system the squared line element $ds^{2}$ in cartesian 
coordinates is:
\andy{krick}
\begin{equation}
ds^{2} = c^{2}dt_{0}^{2}-dx_{0}^{2}-dy_{0}^{2}-dz_{0}^{2}
\label{eq:krick}
\end{equation}
In the formalism of general relativity one has a large liberty in the choice of 
the coordinate system useful for solving a given problem. Since we are doing 
physics and not only differential geometry, once given 
coordinates are choosen, the problem of their interpretation in terms of 
measurable quantities has still to be solved. In the case
of the rotating disk it is simpler to use the coordinates in the right-hand 
side of the following transformations, as done for examples by Langevin 
\cite{lang:21a} and by some textbooks \cite[p. 253]{moll:72a},
\cite[p. 107]{fock:59a}, \cite[p. 281]{lali:59a}.
\andy{adelscott}
\begin{eqnarray}
t_{0}& =& t \nonumber \\
x_{0}& =& r\cos(\varphi +\omega t) \nonumber\\
y_{0}& =& r\sin(\varphi +\omega t) \label{eq:adelscott}\\
z_{0}& =& z\nonumber
\end{eqnarray}
As stressed by M\"{o}ller, in his treatment of the problem of the 
rotating disk, 
the first equation of (\ref{eq:adelscott}) does not mean that we are dealing 
with Newtonian physics, where time dilation effects are missing, but that the 
coordinate $t$ is measured by a clock on the disk, that runs 
$1/\sqrt{1-\omega^{2}r^{2}/c^{2}}$ faster than the clocks of the laboratory, 
when compared at rest. Once this coordinate clock will be put on a disk 
rotating 
with angular velocity $\omega$ at radius $r$, it will have the same rate as the 
clocks in the laboratory, because of the time dilation. 
Similary the second and the third equation of 
(\ref{eq:adelscott}) have a physical meaning, taking account of effects of 
longitudinal 
length contraction, only if the coordinate $\varphi$ is measured with 
tangential coordinates-rods that are $1/\sqrt{1-\omega^{2}r^{2}/c^{2}}$ longer
than the measuring rods of the laboratory when compared at rest.  
Substituting 
(\ref{eq:adelscott}) in (\ref{eq:krick}) one can easily obtain:
\andy{reypnol}
\begin{equation}
ds^{2}=\left(1-\omega^{2}r^{2}/c^{2}\right)\left(cdt\right)^{2}
-2\frac{\omega r^{2}}{c} d\varphi\left(cdt\right)
-dr^{2}-r^{2}d\varphi^{2}-dz^{2}
\label{eq:reypnol}
\end{equation}
Eq. (\ref{eq:reypnol}) defines a metric $g_{ij}$ which is stationary, but not 
static. If $x^{0}=ct$, $x^{1}=r$, $x^{2}=\varphi$, $x^{3}=z$, its element are:
\andy{spaetzli}
\begin{eqnarray}
g_{00}= 1- \omega^{2}r^{2}/c^{2}\;\;;& & g_{11} = g_{33} = -1 \nonumber \\
g_{02} = g_{20} = -\omega r^{2}/c\;\;;& & g_{22} = -r^{2}\;\;, 
\label{eq:spaetzli}
\end{eqnarray}
all other elements being zero. Note that the space-time described by 
(\ref{eq:spaetzli}) is flat because $R_{ijkl}(t_{0},x_{0},y_{0},z_{0}) = 0 
\Rightarrow R_{ijkl}(t,r,\varphi,z) = 0\;\; [i,j,k,l=0,1,2,3]$, 
where $R_{ijkl}$ 
is the Riemann tensor. For the same reason of covariance, the metric defined 
in (\ref{eq:spaetzli}) is necessarly a solution of Einsteins equations in 
empty space 
$R_{ij} =0\;\;[i,j=0,1,2,3]$, where $R_{ij}$ is the Ricci tensor.

\noindent As is well known, in the case of a non-static space-time metric,
the spatial part of the metric is not only given by the 
space-space coefficients of the four dimensional metric, but by:
\andy{mescaline}
\begin{equation}
dl^{2}=\left(-g_{\alpha\beta}+\frac{g_{0\alpha}g_{0\beta}}{g_{00}}\right)
dx^{\alpha}dx^{\beta}
=dr^{2}+dz^{2}+\frac{r^{2}d\varphi^{2}}{1-\omega^{2}r^{2}/c^{2}}\;\;,
\label{eq:mescaline}
\end{equation}
where $0$ is the time index, and $\alpha,\beta$ represent the space indices 
and can take the values $1,2,3$. The right-hand-side of (\ref{eq:mescaline}), 
with $dz=0$, is interpreted by most textbooks as
the standard result, showing that the spatial part of the metric is not 
flat on the rotating disk. The opinion of the author is that it demonstrates 
exactly the contrary, namely that space is flat on the rotating disk. We have 
seen that the angle $\varphi$ was measured with tangeantial coordinates-rods, 
which, at rest, 
are longer than those of the laboratory, the radial measuring rods on the disk 
having the same rest length as those of the laboratory. If in 
contrary we had choosen tangential unit rods which, at rest, have the same 
length as the rods of the laboratory the coordinate $\varphi'$ measured with 
these rods on the disk would satisfy the following equation: 
\andy{kro}
\begin{equation}
d\varphi =d\varphi' \sqrt{1-\omega^{2}r^{2}/c^{2}}\;\;,
\label{eq:kro}
\end{equation}
so that measured with these ``normal'' rods (\ref{eq:mescaline}) becomes:
\andy{dreher}
\begin{equation}
dl^{2}=dr^{2}+dz^{2}+r^{2}d\varphi'^{2}\;\;,
\label{eq:dreher}
\end{equation}
which is a flat metric in polar coordinates.

\noindent We now try to set clocks with an Einstein's synchronisation on 
the disk. Generally, if we send 
a light signal from point $A$ with coordinates $x^{\alpha},\;\alpha=1,2,3$ 
to an infinitesimally near point $B$ with coordinates 
$x^{\alpha}+dx^{\alpha},\;\alpha=1,2,3$ and back, the coordinate time 
difference $dt_{1}$ ($dt_{2}$) for the ``there'' (back) trip is 
obtained by 
solving the equation $ds^{2}=0$. We obtain:
\andy{datura}
\begin{equation}  
dt_{1,2}  =  \frac{1}{cg_{00}}\left[\mp g_{0\alpha}dx^{\alpha}+\sqrt{\left(
g_{0\alpha}g_{0\beta}-g_{\alpha\beta}g_{00}\right)dx^{\alpha}dx^{\beta}}\right]
 =  \pm\frac{\omega r^{2} d\varphi}{c^{2}\left(
1-\omega^{2}r^{2}/c^{2}\right)}
    +\frac{dl_{AB}}{c \sqrt{1-\omega^{2}r^{2}/c^{2}}}
\label{eq:datura}
\end{equation}
By definition the time $t_{A}$, at $A$, which is synchronous with the 
arrival time $t_{B}$ at $B$ is the midtime of departure and 
arrival at
$A$. So, two Einstein-synchronous events are not coordinate-time-synchronous 
(see figure 2)
and have a difference $\Delta t$ such that: 
\andy{amphete}
\begin{equation}
t_{B}=t_{A}+\Delta t = t_{A} 
-\frac{1}{c}\frac{g_{0\alpha}dx^{\alpha}}{g_{00}}=t_{A}+\frac{\omega 
r^{2}d\varphi}{c^{2}\left(1-\omega^{2}r^{2}/c^{2}\right)}
\label{eq:amphete}
\end{equation}
        \vspace{-20pt}
        \begin{figure}[ht]
        \let\picnaturalsize=N
        \def\picsize{7.0cm}
        \def\picfilename{lfig2.eps}
        \ifx\nopictures Y\else{\ifx\epsfloaded Y\else\input epsf \fi
        \let\epsfloaded=Y
        \centerline{\ifx\picnaturalsize N\epsfxsize \picsize\fi
        \epsfbox{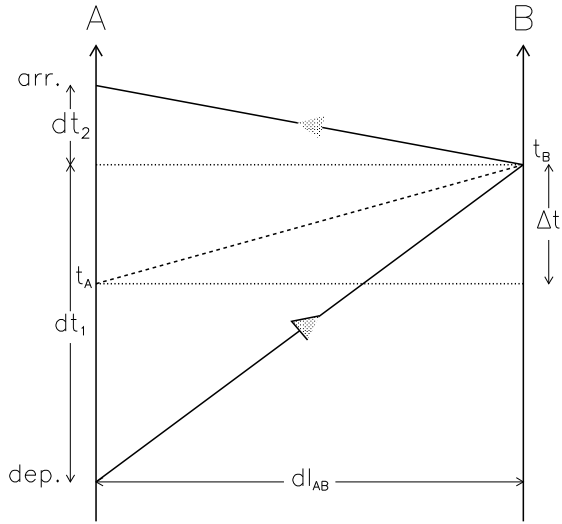}}}\fi
        \vspace{10pt}
        \caption{Illustration of equations (\ref{eq:datura}) and
 (\ref{eq:amphete}). Einstein's synchronous 
events $t_{A}$ and $t_{B}$ are joined with a dashed line. Coordinate-time 
synchronous events are joined with a doted line.}
        \end{figure}
If we generalise this procedure, not only in an infinitesimal domain, but along 
a curve, we obtain that generally it is path dependant, because 
$\Delta t$ is not a total differential in $\varphi$ and $r$. 
As consequence, time 
can not be defined globally on the disk with Einstein's procedure. This is what 
in fact happens on earth: if one synchronises atomic clocks all around the 
earth with Einstein's procedure and comes back to the point of departure 
after a 
whole round trip, a time lag will result. This means that a clock is not 
synchronisable with itself, which is clearly absurd. Moreover, we will 
generally not obtain the same result, when synchronising clock A with clock B 
using two different paths. This mean also that if a clock B is synchronised 
with 
A and a clock C is synchronised with B, C will generally not be synchronised 
with A.  
Hence the physicist Ashby from 
Boulder has said: 
``Thus one discards Einstein's synchronisation in the rotating 
frame''\cite{ashb:93a}. 

\noindent The existence of synchronisation problem is physically strange, 
because if the 
whole disk is initially at rest in the laboratory, with clocks near its rim 
synchronised with the Einstein's procedure, since we are in $\Sigma$, then when 
the disks moves accelerates and attains a constant angular velocity, the clocks 
must slow their rate, but not desynchronise, for symmetry reasons, since they 
had at all time the same speed. From such a point of view, it is difficult to 
see why there should be any difficulty in defining the time on the rotating 
platform. In fact, we see that we can easily define a global time since the 
coordinate time $t$ is already global.
Remembering that the coordinate time $t$ is measured with clocks that run 
faster than clocks at rest in $\Sigma$, we define a global 
time $t'=\sqrt{1-\omega^{2}r^{2}/c^{2}}\;t$. 
The one-way velocity of light on the rim is 
now given by:
\andy{volubilis}
\begin{equation}
c_{\rightarrow}(\pm)=\frac{dl_{AB}}{dt'_{1,2}}=
\frac{dl_{AB}}{dt_{1,2} \sqrt{1-\omega^{2}r^{2}/c^{2}}}=
\frac{c}{1\pm\omega r/c}\;\;,
\label{eq:volubilis}
\end{equation}
where the last step comes from (\ref{eq:mescaline}) and (\ref{eq:datura}) 
with $dr=dz=0$ and $+$ ($-$) stands for the anticlockwise (clockwise) 
propagation of light.
It means that the velocity of light in the tangential inertial frame is also 
equal to:
$c_{\rightarrow}(\pm)=c/(1\pm\beta)$, with $\beta=\omega r/c$
, corresponding to a parameter $s=0$ of (\ref{eq:lexo}) and an angle
$\Theta$ of $0$ ($\pi$). From here, the Sagnac effect can easily be calculated, 
in the same way as in section 2 and the result is found to be identical.

\noindent So we see that the formalism of general relativity is able to 
describe without contradictions the synchronisation of clocks on the rim of 
a rotating disk, but implies a velocity of light in the comoving inertial frame
tangeantial to the rim which is noninvariant and so contradict the clock 
synchronisation of special relativity.

\section{{\small{\sc Conclusion}}}
In inertial systems the synchronisation of clocks is conventional and a set of 
theories equivalent to special relativity, as regards experimental results, can 
be derived. When extended to accelerated motion, the synchronisation is no 
longer conventional and only the theory using $s=0$ from (\ref{eq:stp}) is 
consistent with the {\em Clock Hypothesis}. This conclusion is confirmed in the 
case of the Sagnac effect. An elementary calculation shows that only $s=0$ 
enable to explain the Sagnac effect on the platform. Since one could say that 
accelerated motions have to be calculated with the formalism of general 
relativity, we have done it and shown that in this case also, only a 
noninvariant velocity of light enables to synchronise clocks globally on the 
platform. So, the clock synchronisation used in special relativity is 
inconsistent with the global definition of time used in general relativity.
Moreover we have shown that, in our view, the geometry of the 
rotating platform is flat.

\section{{\small{\sc Acknowledgement}}}

I want to thank the Physics Departement of Bari University for hospitality.

\vspace{-1.2cm}

\begin{thebibliography}{99}

{\small

\vspace{-0.3cm}

\bibitem{poin:04a}
{\sc H. Poincar\'{e},}
{\em M\'{e}langes, l'\'{e}tat 
actuel et l'avenir de la physique math\'{e}matique}, 
 Bull. Phys. Math. {\bf 28} (1904), 302--325.
\vspace{-3.3mm}

\bibitem{eins:05a}
{\sc A. Einstein,}
{\em Zur Elektrodynamik bewgter 
K\"{o}rper}, Ann. Phys. {\bf 17} (1905), 891--921.
\vspace{-3.3mm}

\bibitem{reic:58a}
{\sc H. Reichenbach,}{\em The Philosophy of Space \& 
Time}, Dover, New-York, (1958).
\vspace{-3.3mm}

\bibitem{winn:70a}
{\sc J.A. Winnie,} {\em Special Relativity without
One-Way Velocity Assumptions}, 
 Phil. Mag. {\bf 37} (1970), 81--99 and 223--238.
\vspace{-3.3mm}

\bibitem{grun:73a}
{\sc A. Gr\"{u}nbaum,}
{\em Philosophical Problems of 
Space and Time}, Reidel, Dodrecht (1973).
\vspace{-3.3mm}

\bibitem{jamm:79a}
{\sc M. Jammer,}
{\em Some Fundamental Problems 
in the Special Theory of Relativity}, 
in: Problems in the Foudation of Physics 
(G. Toraldo di Francia, Ed.), North 
Holland, Amsterdam (1979). 
\vspace{-3.3mm}

\bibitem{mase:77a}
{\sc R. Mansouri} and {\sc R.U. Sexl,}
{\em A Test Theory
of Special Relativity: I. Simultaneity and Clock Synchronisation, II. First 
Order Test, III. Second-Order Test}, 
Gen. Relat. Grav. {\bf 8} (1977), 497--513; {\bf 8} (1977) 515--524; 
{\bf 8} (1977), 809--813.
\vspace{-3.3mm}

\bibitem{sjod:79a}
{\sc T. Sj\"{o}din,} {\em Synchronisation in Special 
Relativity and Related Theories}, Nuovo Cim. {\bf 51B } (1979), 229--245.
\vspace{-3.3mm}

\bibitem{cava:89a}
{\sc G. Cavalleri} and {\sc C. Bernasconi,}
{\em Invariance of Light Speed and 
Nonconservation of Simultaneity of Separate Events in 
Prerelativistic Physics and vice versa in Special Relativity}, Nuovo Cim. 
{\bf 104B} (1989), 545-561.
\vspace{-3.3mm}

\bibitem{unga:91a}
{\sc A.A. Ungar,} 
{\em Formalism to Deal with Reichenbach's Special Theory of 
Relativity}, Found. Phys. {\bf 6} (1991), 691--726.
\vspace{-3.3mm}

\bibitem{vest:91a}
{\sc I. Vetharaniam} and {\sc G.E. Stedman,}
{\em Synchronisation Conventions in Test Theories of Special Relativity},
Found. Phys. Lett. {\bf 4} (1991), 275--281.
\vspace{-3.3mm}

\bibitem{anst:92a}
{\sc R. Anderson} and {\sc G.E. Stedman,}
{\em Distance and the Conventionality of Simultaneity in Special Relativity},
Found. Phys. Lett. {\bf 5} (199), 199--220.
\vspace{-3.3mm}

\bibitem{anst:94a}
{\sc R. Anderson} and {\sc G.E. Stedman,}
{\em Spatial Measures in Special Relativity Do Not Empirically Determine 
Simultaneity Relations: A Reply to Coleman and Kort\'{e}},
Found. Phys. Lett. {\bf 7} (1994), 273--283.
\vspace{-3.3mm}

\bibitem{sell:94a}
{\sc F. Selleri,}  {\em Theories Equivalent to Special Relativity},
in: Frontiers of Fundamental Physics, 
(M. Barone and F. Selleri, Eds), 181--192, Plenum Press, New-York (1994). 
\vspace{-3.3mm}

\bibitem{vest:93a}
{\sc I. Vetharaniam}  and {\sc G.E. Stedman,} 
{\em Significance of Precision Tests of Special Relativity}, 
Phys. Lett. A. {\bf 183} (1993), 349--354.
\vspace{-3.3mm}

\bibitem{bail:77a}
{\sc J. Bailey} et al.,
{\em Measurement of Relativistic Time Dilatation for Positive and Negative 
Muons in Circular Orbits}, 
Nature {\bf 268 } (1977), 301--304.
\vspace{-3.3mm}

\bibitem{sell:96a}
{\sc F. Selleri,}
{\em Noninvariant One-Way Velocity of Light},
Found. Phys. {\bf 26} (1996), 641--664.
\vspace{-3.3mm}

\bibitem{mast:93a}
{\sc S.R. Mainwaring} and {\sc G.E. Stedman},
{\em Accelerated Clock Principle in Special Relativity},
Phys. Rev. A {\bf 47} (1993), 3611--3619.
\vspace{-3.3mm}

\bibitem{eisl:87a}
{\sc A. M. Eisele,}
{\em On the Behaviour of an Accelerated Clock,}
Helv. Phys. Act. {\bf 60} (1987), 1024--1037.
\vspace{-3.3mm}

\bibitem{fgoy:96a}
{\sc F. Goy},
{\em Aberration and the Question of Equivalence of Some Ether Theories to 
Special Relativity},
Found. Phys. Lett. {\bf 9} (1996), 165--174.
\vspace{-3.3mm}

\bibitem{sagn:13a}
{\sc G. Sagnac,}
{\em L'\'{e}ther lumineux 
d\'{e}montr\'{e} par l'effet du vent relatif d'\'{e}ther dans un 
interf\'{e}rom\`{e}tre en rotation uniforme}, Comptes Rendus {\bf 157} (1913), 
708--710; ibid., {\em Sur la preuve de la 
r\'{e}alit\'{e} de l'\'{e}ther lumineux par l'exp\'{e}rience de 
l'interf\'{e}rographe tournant}, 1410--1413.
\vspace{-3.3mm}

\bibitem{miga:25a}
{\sc A.A Michelson} and {\sc H.G. Gale,} 
{\em The Effect of Earth Rotation on the Velocity of Light}, 
Astro. J. {\bf 61} (1925), 137--145.
\vspace{-3.3mm}

\bibitem{anbi:94a}
{\sc R. Anderson, H.R. Bilger} and {\sc G.E. Stedman,}
{\em ``Sagnac'' effect: A Century of Earth Rotated Interferometers},
Am. J. Phys. {\bf 62} (1994), 975--985.
\vspace{-3.3mm}

\bibitem{hani:93a}
{\sc H. Hasselbach} and {\sc M. Nicklaus},
{\em Sagnac Experiment with Electrons: Observation of the Rotational Phase 
Shift of Electron Waves in Vacuum} 
Phys. Rev. A {\bf 48} (1993), 143--151.
\vspace{-3.3mm}

\bibitem{coov:75a}
{\sc R. Colella, A.W. Overhauser, J.L. Staudenmann} and {\sc S.A. Werner,} 
{\em Gravity and Inertia in Quantum Mechanics},
Phys. Rev. A {\bf 21 } (1980), 1419--1438.
\vspace{-3.3mm}

\bibitem{stco:94a}
{\sc P. Storey} and {\sc C. Cohen-Tannoudji,}
{\em The 
Feynman Path Integral Approach to Atomic Interferometry. A Tutorial}, 
J. Phys. II France {\bf 4} (1994), 1999--2027.
\vspace{-3.3mm}

\bibitem{fach:96a}
{\sc D. Fargion, L. Chiatti} and {\sc A. Aiello},
{\em Quantum Mach Effect by Sagnac Phase Shift on Cooper Pairs in rf-SQUID},
preprint astro-ph/9606117, babbage.sissa.it (1996), 9 pages.
\vspace{-3.3mm}

\bibitem{alal:85a}
{\sc D.W. Allan} et al.,
{\em Accuracy of 
International Time and Frequency Comparisons Via Global Positioning System 
Satellites in Common-View}, IEEE Trans. Instr. Meas. 
{\bf IM-34} (1985), 118--125;  Science, {\em Around-the-World
Relativistic Sagnac Experiment} {\bf 228} (1985), 69--70.
\vspace{-3.3mm}

\bibitem{lang:21a}
{\sc P. Langevin,}
{\em Sur la th\'{e}orie de la 
relativit\'{e} et l'exp\'{e}rience de M. Sagnac}, Comptes Rendus {\bf 
173} (1921), 831--834; ibid., {\em Sur 
l'exp\'{e}rience de Sagnac}, {\bf 205} (1937), 304--306.
\vspace{-3.3mm}

\bibitem{anan:81a}
{\sc J. Anandan,}
{\em Sagnac Effect in Relativistic and Nonrelativistic Physics}, Phys. Rev. D 
{\bf 24} (1981), 338--346.
\vspace{-3.3mm}

\bibitem{dini:90a}
{\sc D. Dieks} and {\sc G. Nienhuis,}
{\em Relativistic 
Aspects of Nonrelativisic Quantum Mechanics}, Am. J. Phys. {\bf 58}
(1990), 650--655.
\vspace{-3.3mm}

\bibitem{post:67a}
{\sc E.J. Post,}
{\em Sagnac Effect},  Rev. Mod. Phys. {\bf 39} (1967), 475--493.
\vspace{-3.3mm}

\bibitem{moll:72a}
{\sc C. M\"{o}ller,}
{\em The Theory of Relativity}, second edition, Clarendon Press, Oxford (1972).
\vspace{-3.3mm} 

\bibitem{fock:59a}
{\sc V. Fock,}
{\em The Theory of Space Time and Gravitation,}
Pergamon Press, London (1959).
\vspace{-3.3mm}

\bibitem{lali:59a}
{\sc L. Landau} and {\sc E.M. Lifschitz,}
{\em The Classical Theory of Fields,} 
Pergamon Press, London (1959).
\vspace{-3.3mm}

\bibitem{ashb:93a} {\sc N. Ashby,}
{\em Relativity in the Future of Engineering}, Proc. of the IEEE 
International Frequency Control Symposium (1993), 2--14.
\vspace{-3.3mm}

}

\end{thebibliography}

\end{document}